\documentclass[epjST]{svjour}
\usepackage{graphics}
\usepackage{multirow}
\begin{document}
\title{Lossless Data Compression with Error Detection using Cantor Set}
\author{Nithin Nagaraj\inst{1,2}\fnmsep\thanks{\email{nithin.nagaraj@gmail.com}}}
\institute{Department of Electronics and Communication Engineering, Amrita School of Engineering, Amrita Vishwa Vidyapeetham, Amritapuri Campus, Kollam, Kerala 690525. \and Adjunct Faculty, School of Natural Sciences and Engineering, National Institute of Advanced Studies, Indian Institute of Science Campus, Bangalore 560012.
}
\abstract{
In 2009, a lossless compression algorithm based on 1D chaotic maps known as Generalized Lur\"{o}th Series (or GLS) has been proposed. This algorithm (GLS-coding) encodes the input message as a symbolic sequence on an appropriate 1D chaotic map (GLS) and the compressed file is obtained as the initial value by iterating backwards on the map. For ergodic sources, it was shown that GLS-coding achieves the best possible lossless compression (in the noiseless setting) bounded by Shannon entropy. However, in the presence of noise, even small errors in the compressed file leads to catastrophic decoding errors owing to sensitive dependence on initial values. In this paper, we first show that Repetition codes $\mathcal{R}_n$ (every symbol is repeated $n$ times, where $n$ is a positive odd integer), the oldest and the most basic error correction and detection codes in literature, actually lie on a Cantor set with a fractal dimension of $\frac{1}{n}$, which is also the rate of the code. Inspired by this, we incorporate error detection capability to GLS-coding by ensuring that the compressed file (initial value on the map) lies on a Cantor set of measure zero.  Even a 1-bit error in the initial value will throw it outside the Cantor set which can be detected while decoding. The error detection performance (and also the rate of the code) can be controlled by the fractal dimension of the Cantor set and could be suitably adjusted depending on the noise level of the communication channel.
}
\maketitle
\section{Introduction}
\label{intro}
Lossless data compression is a fundamental block in most storage and communication systems~\cite{Bose}. We consider an i.i.d binary message source $S$ which emits binary symbols (`0' and `1') with probability of `0' being $p$ ($0 < p < 1$). Consider a binary message $MSG$ of length $N$ from such a source. Shannon~\cite{Shannon48} showed that such a message can be losslessly compressed to $\geq H(S)\cdot N$ bits where $H(S)$ is the Shannon's entropy of the source. The value $H(S)\cdot N$ is the {\it best} possible lossless compression. For an individual binary message $MSG$ from such an i.i.d source, we can compute $H(\cdot)$ as follows:
\begin{equation}
H(MSG) = -p\log_2(p) -(1-p)\log_2(1-p)~~~~bits/symbol,
\end{equation}
where $p =\frac{\#~of~zeros~in~MSG}{N}$. Thus $H(MSG) \cdot N$ is the ultimate limit of lossless data compression that can be achieved.  There are several lossless compression algorithms in literature -  Shannon-Fano coding, Huffman coding, Arithmetic coding, Lempel-Ziv coding and others~\cite{Cover,Sayood}. Among these, Arithmetic coding~\cite{Rissanen} achieves the Shannon entropy limit for increasing message lengths. Arithmetic coding is used extensively in several practical applications owing to its speed and efficiency. In fact, it is used in JPEG2000~\cite{jpeg2000}, the international standard for still image compression, replacing the popular Huffman coding which was used in  the earlier JPEG~\cite{jpeg} standard.

In 2009~\cite{NithinGLS}, it was shown that Arithmetic coding is closely related to a 1D non-linear chaotic dynamical system known as Generalized Lur\"{o}th Series (GLS). Specifically, it was shown that lossless compression or encoding of the binary $MSG$ can be performed as follows. First, the $MSG$ is treated as a symbolic sequence on an {\it appropriately} chosen GLS. The initial value on the GLS corresponding to this symbolic sequence is computed by iterating backwards. This initial value (written in binary) serves as the compressed file. For decompression (or decoding), we start with this initial value (the compressed file) and iterate forwards on the (same) GLS, and record the symbolic sequence. This symbolic sequence is the decoded $MSG$.  Such a simple lossless compression algorithm (known as GLS-coding) was proved to achieve Shannon's entropy limit~\cite{NithinGLS}. Arithmetic coding turns out to be a special case of GLS-coding~\cite{NithinGLS}.

Unfortunately, it turns out that Arithmetic coding (as well as GLS-coding) is very sensitive to errors. Even a single bit error in the compressed file can lead to catastrophic decoding errors. This has been well documented in the data compression literature~\cite{Boyd1997} and researchers have since been trying to enhance Arithmetic coding with error detection and correction properties~\cite{ac1,ac2}. In this work, our aim is to incorporate error detection into GLS-coding by using Cantor set while not sacrificing much on the lossless compression ratio performance of GLS-coding. As we shall demonstrate, Cantor set has desirable properties that enable efficient error detection while GLS-decoding.

The paper is organized as follows. In section 2, we briefly describe GLS-coding of a binary message. In the same section, we show how a single bit error in the GLS-coded file can lead to catastrophic decoding errors. In section 3, we explore Cantor set and describe its wonderful properties that enable error detection. Repetition codes, a very popular classical error detection and correction code is shown to be a Cantor set. Inspired by this, we incorporate error detection into GLS-coding using a Cantor set in section 4. Extensive simulation results are also provided in the same section to demonstrate the efficiency of our method. We conclude with open challenging problems in section 5.
\section{GLS-Coding and Decoding}
\label{gls-coding}
\begin{figure}
\begin{center}
\centering
\resizebox{0.5\columnwidth}{!}{
\includegraphics{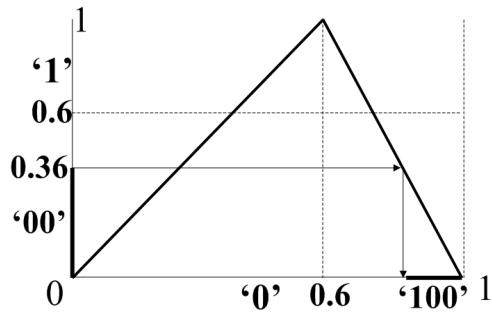}}
\caption{GLS-coding: An example. For the $MSG = 0010110100$, the above figure shows the backward-iteration from $00$ to $100$. This process is repeated until we have a final interval $[START, END)$ for the entire $MSG$. This means that all initial values in $[START, END)$ have the symbolic sequence ($=MSG$) on the above map. The initial value $x_0 = \frac{START+END}{2}$ in binary is written as the output compressed file.} \label{fig:glscoding}
\end{center}
\end{figure}

In this section, we shall briefly describe GLS-Coding proposed in~\cite{NithinGLS}. We are given a binary message $MSG = 0010110100$ (say) from an i.i.d source $S$. Our aim is to losslesly compress $MSG$. To this end, we first determine the probability of zeros in $MSG$, which is $p=0.6$. The skew-tent\footnote{One could alternately use the skew-binary map. There are eight possible {\it modes} of GLS in this case and all are equivalent in terms of compression ratio performance. See~\cite{NithinGLS} for further details.}map $T$ is constructed with two intervals: $[0, 0.6)$ for the symbol zero and $[0.6 , 1)$ for the symbol one. We treat the given $MSG$ as a symbolic sequence on $T$ and determine (by iterating backwards on the map $T$, see Fig.~\ref{fig:glscoding}) the final interval  $[START, END)$ in which all initial conditions have the same symbolic sequence ($=MSG$). We take the mid-point $x_0 = \frac{START+END}{2}$ as the compressed file. Since $x_0$ is a real number (between 0 and 1), we write its binary representation to the file. The number of bits needed to represent the compressed file $x_0$ is  $ \lceil -\log_2(END-START) \rceil $ bits. This is proved to be optimal (best possible compression) in~\cite{NithinGLS} and Arithmetic coding is shown to be a special case of GLS-coding.

GLS-Decoding is straightforward. At the decoder, given the value of $p$, we construct the skew-tent map $T$ as before. Given that we know $x_0$, all that we need to do is iterate forwards on the map $T$ for $N$ ($=$ length of $MSG$) iterations and output the symbolic sequence (if the iterate falls in $[0, 0.6)$ output symbol `0', else output `1'). This is the decoded message and in the absence of any noise, this is exactly the same as $MSG$ which was input to the encoder.
\subsection{Effect of single-bit error on GLS-decoding: Sensitive dependence on initial value of the map}
Thus far, we have discussed lossless coding in the absence of any
kind of noise. However, in practical applications, noise gets added
invariably. For example, in storage devices like compact discs, a
simple scratch could result in flipping some bits which are ones to
zeroes and zeroes to ones. In applications involving transmission of
data from one location to another, noise gets added in the
communication channel. Sources of noise could include signals from
other sources which interfere, thermal noise and noise from physical
sources. Wireless channels are known to be extremely
noisy~\cite{WirelessCommBook}. Satellite communications also suffer
from noise. Noise introduces error in the data. If the data is in a
compressed form, then it is quite likely that the decoder would be
unable to decode or would decode incorrectly. Depending on the type
of compression that is employed, the entire data can be rendered
non-decodable due to the corruption of only a few bits of the
compressed file.

\begin{figure}
\begin{center}
\centering
\resizebox{1.0\columnwidth}{!}{
\includegraphics{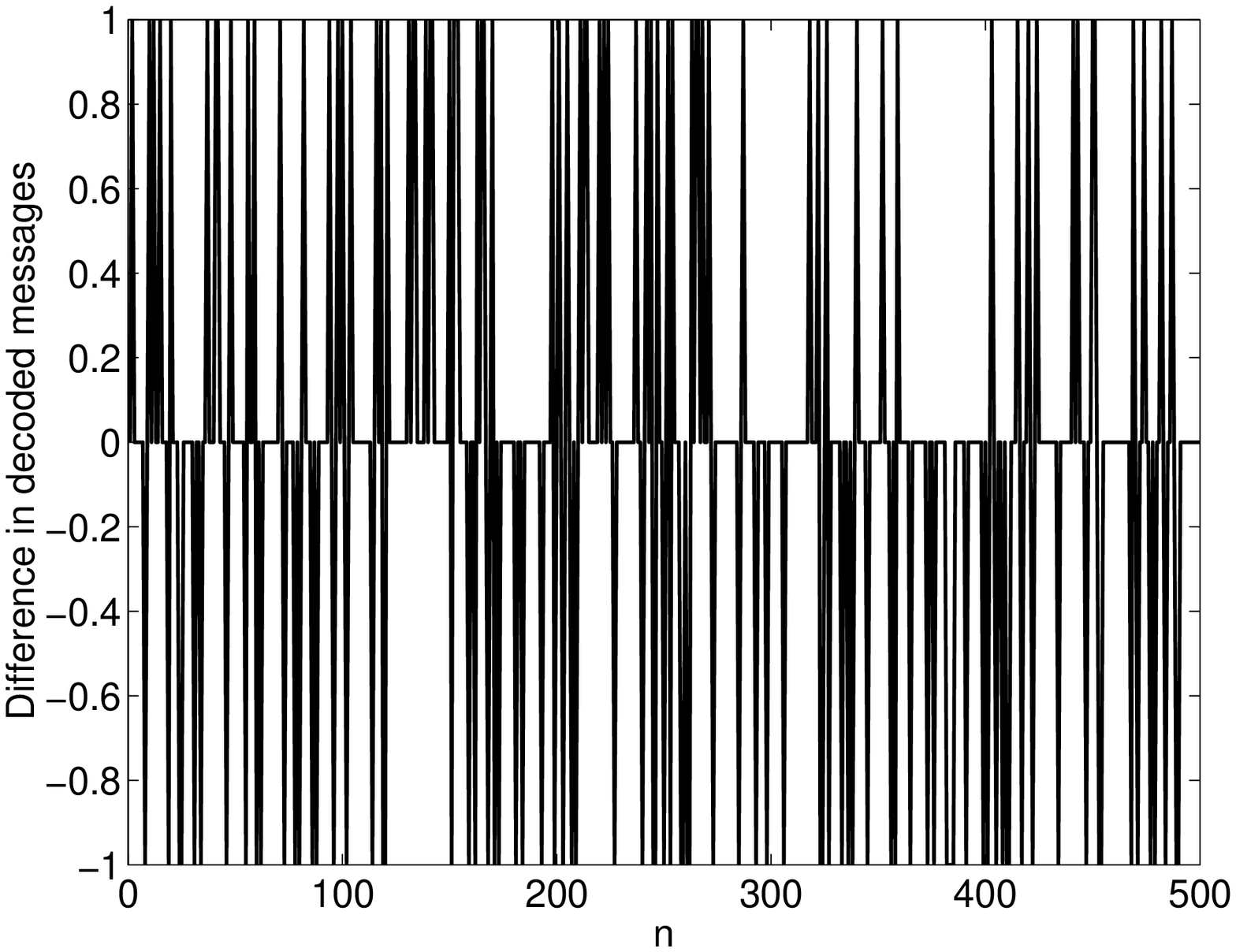} \hspace{0.25in}
\includegraphics{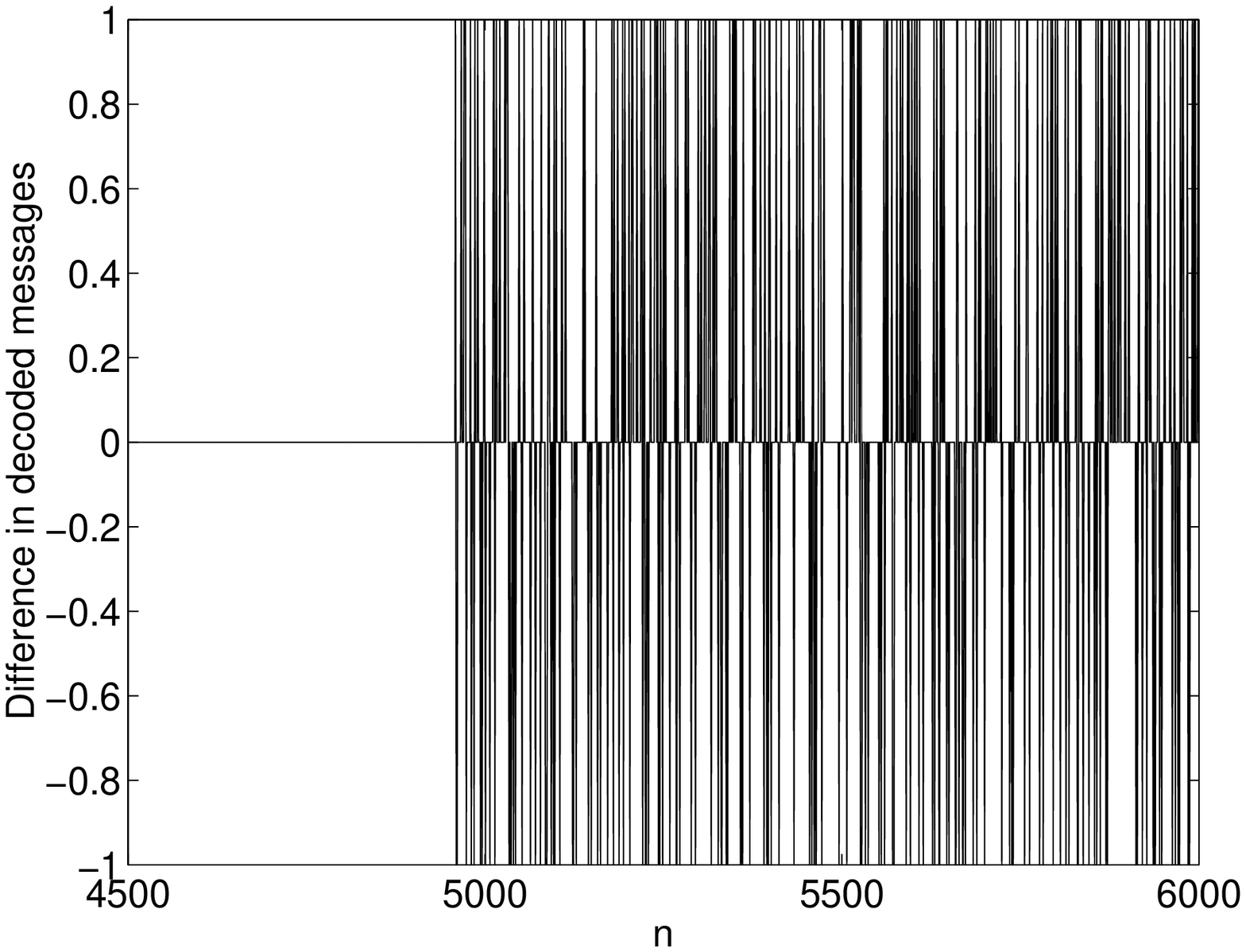}
}
\caption{Effect of noise on
GLS-coding: (a) Left: The first bit of the compressed file is flipped. The
decoded message is very different from the actual intended message. (b) Right: The
middle bit (bit no. 3610) of the compressed file is flipped. The
first 5000 bits are decoded without error and the remaining 5000
bits show lots of decoding erro. ($p=0.2, N=10,000$, Compressed file
size $= 7220$ bits). In both cases, only part of the difference is
shown.} \label{fig:figeffectofnoise}
\end{center}
\end{figure}

In GLS-coding, the compressed file is the {\it initial value}
of the symbolic sequence (the message $MSG$) on the appropriate GLS. Since
GLS is a chaotic map, it exhibits {\it sensitive dependence on
initial values}, the hallmark of deterministic chaos. A small
perturbation in the {\it initial value} will result in a
symbolic sequence which is uncorrelated to the original symbolic
sequence after a few iterations. This means that with a very high probability, even a slight amount of noise that is
added to the initial value (compressed file) will result in a wrongly
decoded message (symbolic sequence), which will be very different from
the actual intended message. This is demonstrated in
Fig.~\ref{fig:figeffectofnoise}. The first bit of the compressed
file is flipped and GLS-decoding is performed. The difference in the
decoded message from the original message is shown in
Fig.~\ref{fig:figeffectofnoise}(a). As it can be seen, the decoded
message is very different from the original message. On the other
hand, if the middle bit of the compressed file is flipped then the
decoded image is accurate up to 5000 bits and the remaining 5000
bits are wrongly decoded (Fig.~\ref{fig:figeffectofnoise}(b)). The
error affects only those bits which are subsequently decoded.
\section{Error Detection using Cantor Set}
\label{cantor}
In GLS-coding, every real number on the interval $[0,1)$ represents an
initial value (compressed file). Thus, any error in the initial
value will result in another real number which is also an initial value, but for an entirely different symbolic sequence (message). It represents a valid compressed file which
decodes to a different message. Therefore in order to detect errors, we necessarily require that
when noise gets added to the initial value while transmission on the communication channel, it should result in a value
that is not a valid compressed file, so that at the decoder it can be flagged for error. This necessarily implies that
not all real numbers in the interval $[0,1)$ can be valid compressed
files. We need to restrict the set of valid compressed files to a
small subset of $[0,1)$. This subset should be uncountable and dense
since it should be able to decode into all possible (infinite
length) messages. At the same time, it should have negligible
measure (zero measure) so that when noise is added, the probability
that it falls outside the set is 1. Cantor sets provide the perfect solution.

\subsection{The Cantor Set}
The well known middle-third Cantor set~\cite{Strogatz1994} is a good
example to illustrate this idea. All real numbers between 0 and 1
which do not have 1 in their ternary expansion belong to this Cantor
set (call it $C$). We note down the following ``paradoxical''
aspects of Cantor sets as observed in \cite{Strogatz1994}:

\begin{enumerate}
\item Cantor set $C$ is ``totally disconnected''. This means that
$C$ contains only single points and no intervals. In this sense, all
points in $C$ are well separated from each other.

\item On the other hand, $C$ contains no ``isolated points''. This
means that every point in $C$ has a neighbor arbitrarily close by.
\end{enumerate}

These two ``paradoxical'' aspects of Cantor sets (not just for the
middle third Cantor set, but even for topological Cantor
sets\footnote{Topological Cantor sets are not self-similar.}) are
actually very beneficial for error detection and correction.
Property 1 implies that a small error will ensure that the resulting
point is not in $C$ while Property 2 ensures that we can always find
the nearest point in $C$ that can be decoded. Self-similar Cantor
sets are fractal (their dimension is not an integer).

We shall show that repetition codes, one of the oldest error
detection/correction codes lie on a Cantor set.

\subsection{Repetition Codes $\mathcal{R}_n$ lie on a Cantor Set}
Repetition codes are the oldest and most basic error detection and
correction codes in coding theory. They are frequently used in
applications where the cost and complexity of encoding and decoding
are a primary concern. Pavel Loskot et al.~\cite{RepetitionCode1}
provide a long list of practical applications of repetition codes. Repetition codes
are robust against impulsive noise and used in retransmission
protocols, spread spectrum systems, multicarrier systems, infrared
communications, transmit delay diversity, BLAST signaling,
rate-matching in cellular systems, and synchronization of
ultrawideband systems \footnote{For further details, please see the
references in~\cite{RepetitionCode1}.}. Thus, repetition codes are
very useful in communications.

They are described as follows. Consider a message from a binary
alphabet $\{ 0, 1\}$. A repetition code $\mathcal{R}_n$ is a block
code which assigns:
\begin{eqnarray*}
0 & \mapsto & \underbrace{0 \ldots 0 }_n \\
1 & \mapsto & \underbrace{1 \ldots 1 }_n .
\end{eqnarray*}
 Thus every symbol of the message is repeated $n$ times
where $n$ is a positive odd integer. For example, the message $M =$
`$01101001$' is coded as $\mathcal{R}_3(M) =$ `$000 111 111 000 111
000 000 111$'. $\mathcal{R}_n$ can correct up to $\frac{n-1}{2}$ bit
errors since the minimum hamming distance of $\mathcal{R}_n$ is $n$
(hamming distance between two binary codewords is the number of
locations in which they differ; minimum hamming distance is an
important characteristic of error correction
codes~\cite{MacKay2003}). A majority count in every block of $n$
symbols acts as a very simple but efficient decoding algorithm. Thus
$\mathcal{R}_3$ can correct all $1$ bit errors and $\mathcal{R}_5$
can correct all $2$ bit and all 1-bit errors. The repetition code $\mathcal{R}_n$ is
a linear block code\footnote{Codes where sum of two codewords modulo
2 is another codeword are known as (binary) linear codes.} with a rate $=
\frac{1}{n}$.

\begin{figure}
\begin{center}
\centering
\resizebox{0.9\columnwidth}{!}{
\includegraphics{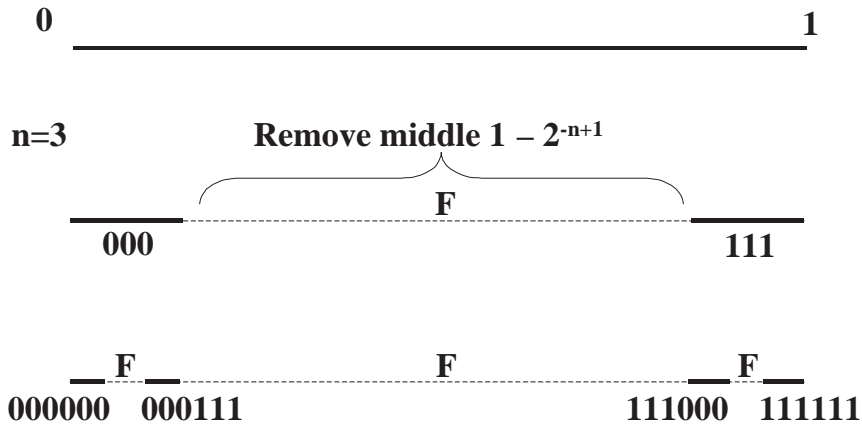}
}
\caption{Repetition codes $\mathcal{R}_n$ lie on a Cantor set:
recursively remove the middle ($1 - 2^{-n+1}$) fraction. As an
example, $\mathcal{R}_3$ is depicted above. The box-counting
dimension of the Cantor set is $D = \frac{1}{n}$. This is equal to
the rate of the code. } \label{fig:repetitioncodes}
\end{center}
\end{figure}

We shall provide a new interpretation of $\mathcal{R}_n$, inspired by Cantor set. Start with
the real line segment $(0,1]$. Remove the middle ($1 - 2^{-n+1}$)
fraction of the set $(0,1]$. In the remaining two intervals, remove
the same fraction and repeat this process in a recursive fashion
(refer to Fig.~\ref{fig:repetitioncodes}). When this process is carried over
an infinite number of times, the set that remains is a Cantor set.
Furthermore, the binary representation of every element of the
Cantor set forms the codewords of $\mathcal{R}_n$.

In order to see this, consider $n=3$. Fig.~\ref{fig:repetitioncodes} shows
how $\mathcal{R}_3$ is recursively constructed. If the above step is
terminated at iteration $k$, then there remains a set of intervals
whose binary expansion (of length $nk$) contains the codewords for
all possible binary messages of length $k$. For example, at $k=2$
for $\mathcal{R}_3$, we can see that there are four intervals which
contains real numbers with binary expansions starting from $000000$,
$000111$, $111000$ and $111111$. These are the codewords for the
messages $00$, $01$, $10$ and $11$ respectively. In the limit of
this process, the set results in a Cantor set of measure zero which
contains codewords for all binary messages which are infinitely
long.

\subsection{Box-Counting Dimension of $\mathcal{R}_n$}
We noted that repetition codes $\mathcal{R}_n$ lie on a Cantor set.
It is very easy to compute the box-counting dimension of this Cantor
set: $D= \lim_{\delta \rightarrow 0} \frac{log N(\delta)}{log (1 /
\delta)}$ where $N(\delta)$ is the number of boxes of size $\delta$
needed to cover the set. For $\mathcal{R}_n$, the box-counting
dimension $D = \frac{1}{n}$ which is equal to the rate of the code.
This establishes a very important connection between the properties
of the Cantor set and the property of the code. The rate of the code
effectively conveys the amount of information bits which are
transmitted for every bit of the code. In general, a high rate code is desirable, though typically the error detection/correction performance and the rate are inversely related (in the sense that it is very difficult to design a high rate code which also has very good error detection/correction performance).

\section{Incorporating Error Detection into GLS-coding using a Cantor Set}
Repetition codes which are error detection/correction codes lie on a
Cantor set. How can we extend this idea of placing codewords on a
Cantor set for GLS-coding? Here, we establish the connection between
repetition codes and GLS-coding.

\subsection{Repetition Codes Re-visited}
It is a common view to look at Repetition codes as block codes\footnote{A block code is one where every finite block of input symbols of size $k$ is mapped to a unique codeword of $n$ symbols with $n>k$.}. In
this section, we view them as GLS-coding with a forbidden symbol.

We could re-interpret Fig.~\ref{fig:repetitioncodes} in a different way. Let
the middle $1 - 2^{-n+1}$ interval be reserved for the forbidden
symbol `$F$' (this symbol never occurs in the message to be encoded)
and the intervals $[0, 2^{-n})$ and $[1-2^{-n}, 1)$ correspond to
the symbols `0' and `1' respectively. We treat all binary messages
as symbolic sequences on this modified map and perform GLS-coding,
i.e. find the initial value of a given message $M$. For
GLS-coding, we are treating the alphabet $\{ 0, F, 1 \}$ as taking
the probabilities $\{ 2^{-n}, 1 - 2^{-n+1}, 2^{-n} \}$ respectively.
The resulting initial value of GLS-coding is the codeword for the message and
it turns out that it is the same as $\mathcal{R}_n(M)$. Thus we have
interpreted $\mathcal{R}_n(M)$ as a joint source channel code where
the source has three alphabets and we are encoding messages that
contain only 0 and 1.  

By reserving a forbidden symbol `$F$' which is not used in encoding,
all pre-images of the interval corresponding to `$F$' have to be
removed. Thus, we have effectively created the same Cantor set that
was referred to in the previous section. For error detection, one
has to start with the initial value and iterate forwards on the
modified map and record the symbolic sequence. If the symbolic
sequence while decoding contains the symbol `$F$', then it invariably means that the initial value is
not a part of the Cantor set and hence not a valid codeword of
$\mathcal{R}_n$, thereby detecting that an error has occurred. Thus checking whether the initial value received belongs to the Cantor set or not is used for error detection at the decoder.

\subsection{GLS-coding with a Forbidden Symbol}
We have presented two new ways of looking at Repetition codes - 1)
the codewords of $\mathcal{R}_n$ lie on a Cantor set and 2) coding a
message is the same as performing GLS-coding with a forbidden symbol
reserved on the interval $[0,1)$. The two are essentially the same
because, by reserving a forbidden symbol $F$, we have effectively
created a Cantor set on which all the codewords lie. But the fact
that we can view $\mathcal{R}_n$ as GLS-codes enables us to see them
as joint source channel codes for the source with alphabets $\{ 0,
F, 1 \}$ and with probability distribution $\{ 2^{-n}, 1 - 2^{-n+1},
2^{-n} \}$ respectively. The natural question to ask is whether we
can use the same method for a different probability distribution of
$0$ and $1$. The answer is positive.

Instead of reserving a forbidden symbol $F$ of length $1 -
2^{-n+1}$, we could chose any arbitrary value $\epsilon > 0$ for the
forbidden symbol. The value of $\epsilon$ determines the amount of
redundancy that is going to be available for error
detection/correction. It controls the fractal dimension of the
Cantor set and hence the rate of the code. As $\epsilon$ increases,
error detection/correction property improves at the cost of a slight
reduction in compression ratio (note that the compression is still
lossless, but no longer Shannon optimal). The probability of the
symbol `0' is $p$, but only $(1-\epsilon)p$ is allocated on the
interval [0,1). Similarly, for the symbol `1': $(1-\epsilon)(1-p)$
is allocated. This single parameter $\epsilon$ can be tuned for
trade-off between error control and lossless compression ratio. We shall show
that a very small value of $\epsilon$ is sufficient for detecting
errors without significantly increasing the compressed file size.

For encoding, as before, the binary message is treated as a symbolic
sequence on the modified GLS with the forbidden symbol `$F$' and the
initial value is determined. The initial value which is now
on the Cantor set is the compressed file which is stored and/or
transmitted to the decoder.

\subsection{Error Detection while GLS-Decoding}
The decoder is the same as before except that we now have error
detection capability. If while GLS-decoding, the forbidden symbol `$F$'
is encountered (this can happen only if noise corrupts the initial value/compressed file and throws it outside the Cantor set), then it is declared that an error has been detected. The
decoder can then request the encoder to re-transmit the compressed
file as is done in several protocols~\cite{ARQ}. However, this
scheme does not correct the error. It is quite possible that the noise is
such that the initial value gets modified into another value which also happens to fall inside
the Cantor set, in which case the decoder will not be able to detect
the error (and thus we end up wrongly decoding the message). But, the probability of this occurring is very small (it
is zero in the case of messages having infinite length since the
measure of the Cantor set is zero). For finite length messages, the
probability of such an event is given by the measure of the set of
codewords (which is non-zero). In the following section, we perform rigorous experimental
tests of the proposed approach.

\subsection{Simulation Results}
Three different values of $\epsilon$ ($\epsilon_1 = 0.005$,
$\epsilon_2 = 0.03$, $\epsilon_3 = 0.05$) for the length of the
interval corresponding to the forbidden symbol `$F$' were used. The
amount of redundancy that is added is easy to determine. By
introducing the forbidden symbol of length $\epsilon$, the valid
symbols occupy a sub-interval of length $1-\epsilon$. Thus, each
time a symbol with probability $p$ is encoded,
$-\log_2((1-\epsilon)p)$ bits will be spent, whereas only
$-\log_2(p)$ bits would have been spent without the forbidden
symbol. Thus, the amount of redundancy is $R(\epsilon) =
-\log_2((1-\epsilon)p) + \log_2(p) = -\log_2(1-\epsilon)$
bits/symbol. For $N$ symbols, this would be $N \cdot R(\epsilon)$
bits rounded to the nearest highest integer. Thus the rate of the
code will be:

\begin{equation}
\label{eqn:1} \textrm{Rate~} = \frac{1}{1+R(\epsilon)} = \frac{1}{1
- \log_2(1-\epsilon)}.
\end{equation}

As expected, this is equal to the box-counting dimension of the
Cantor set. Thus, by plugging in $\epsilon = 1 - 2^{-n+1}$ in
Equation~\ref{eqn:1}, we obtain the rate of the repetition codes as
$\frac{1}{n}$.

We introduced a single bit error (one bit is flipped in the entire
compressed file) towards the end of the compressed file for binary
i.i.d sources ($p=0.1, 0.3$). Note that, it is much more difficult
to detect errors if they happen towards the end of the compressed
file than if it occurred in the beginning of the file. This is
because, any error can only affect decoding for subsequent bits and
if the error was towards the end-of-file (EoF), not many bits are
available to catch it. The location of the single bit error was
varied from the last bit to the 250th bit from end-of-file. This
way, we can test the proposed method under the worst condition.

\begin{table}[!h]
\centering
\caption[GLS-Coding with Forbidden Symbol: Redundancy]{GLS-Coding with Forbidden Symbol: Redundancy.} %
\label{tab:table_GLSFred} %
\vspace{0.1in}
\begin{tabular}{|@{}c@{}|c|@{}c@{}|@{}c||c@{}|@{}c||c@{}|@{}c||c@{}|}
  \hline
   &  & Compressed & \multicolumn{2}{|c|}{$\epsilon_1 = 0.005$}   & \multicolumn{2}{|c|}{$\epsilon_2  = 0.03$} & \multicolumn{2}{|c|}{$\epsilon_3  = 0.05$}\\
   \cline{4-5} \cline{6-7} \cline{8-9}
  $N$    & $p$    & File Size  & $N \cdot R(\epsilon_1)$    & File Size & $N \cdot R(\epsilon_2)$  & File Size & $N \cdot R(\epsilon_3)$  &  File Size\\

      &      & (bits)     & (bits)                     & (bits)      &      (bits) &      (bits) &      (bits) &      (bits)\\
  \hline
  10000 & 0.1 & 4690     & 72 & 4762 & 440 & 5130 & 740 & 5430\\
  10000 & 0.3 & 8812     & 72 & 8881 & 440 & 9253 & 740 & 9552\\
  \hline
\end{tabular}
\end{table}
\begin{table}[!h]
\centering
\caption[GLS-Coding with Forbidden Symbol: Error Detection ($p=0.1$)]{GLS-Coding with Forbidden Symbol: Error Detection ($p=0.1$).} %
\label{tab:table_GLSErrorP1} %
\vspace{0.1in}
\begin{tabular}{|c|@{}c@{}|@{}c@{}|@{}c@{}|@{}c@{}|@{}c@{}|@{}c@{}|}
  \hline
  Distance of      &   \multicolumn{2}{|c|}{ }   &   \multicolumn{2}{|c|}{ }  & \multicolumn{2}{|c|}{ } \\
  single bit-error & \multicolumn{2}{|c|}{$\epsilon_1 = 0.005$} & \multicolumn{2}{|c|}{$\epsilon_2 = 0.03$} & \multicolumn{2}{|c|}{$\epsilon_3 = 0.05$}\\
  \cline{2-3}   \cline{4-5} \cline{6-7}
  from EoF         & Detected & Undetected & Detected & Undetected & Detected & Undetected \\
  \cline{1-7}
  1-50             & 16        & 34          & 31         & 19    & 41   & 9 \\
  51-100           & 18        & 32          & 48         & 2     & 49   & 1 \\
  101-150          & 32       & 18          & 49         & 1      & 50   & 0\\
  151-250          & 92        & 8          & 100         & 0     & 100  & 0\\
   \cline{1-7}
  Total            & 158        & 92          & 228         & 22  & 240  & 10\\
  \hline
\end{tabular}
\end{table}

\begin{table}[!h]
\centering
\caption[GLS-Coding with Forbidden Symbol: Error Detection ($p=0.3$)]{GLS-Coding with Forbidden Symbol: Error Detection ($p=0.3$).} %
\label{tab:table_GLSErrorP2} %
\vspace{0.1in}
\begin{tabular}{|c|@{}c@{}|@{}c@{}|@{}c@{}|@{}c@{}|@{}c@{}|@{}c@{}|}
  \hline
  Distance of      &   \multicolumn{2}{|c|}{ }   &   \multicolumn{2}{|c|}{ }  & \multicolumn{2}{|c|}{ } \\
  single bit-error & \multicolumn{2}{|c|}{$\epsilon_1 = 0.005$} & \multicolumn{2}{|c|}{$\epsilon_2 = 0.03$} & \multicolumn{2}{|c|}{$\epsilon_3 = 0.05$}\\
  \cline{2-3}   \cline{4-5} \cline{6-7}
  from EoF         & Detected & Undetected & Detected & Undetected & Detected & Undetected \\
  \cline{1-7}
  1-50             & 9        & 41          & 27         & 23      & 36 & 14       \\
  51-100           & 15       & 35          & 43         & 7       & 48 & 2        \\
  101-150          & 29       & 21          & 49         & 1       & 50 & 0        \\
  151-250          & 69        & 31          & 99        & 1       & 100 & 0       \\
   \cline{1-7}
  Total            & 122       & 128          & 218         & 32   & 234 & 16       \\
  \hline
\end{tabular}
\end{table}

\begin{table}[!h]
\centering
\caption[GLS-Coding with Forbidden Symbol: Efficiency]{GLS-Coding with Forbidden Symbol: Efficiency.} %
\label{tab:table_GLSErrorFinal} %
\vspace{0.1in}
\begin{tabular}{|c|c|c|c|}
  \hline
  $p$ &  \multicolumn{3}{|c|}{\% of errors detected }   \\
  \cline{2-4}
   & $\epsilon_1 = 0.005$ & $\epsilon_2 = 0.03$ & $\epsilon_3 = 0.05$\\
   \hline
  0.1 & 63.2 \%& 91.2 \%  & 96.0 \% \\
  0.3 & 48.8 \%& 87.2 \%  & 93.6 \% \\
  \hline
\end{tabular}
\end{table}

Table~\ref{tab:table_GLSFred} shows the amount of redundancy owing
to the allocation of the forbidden symbol.
Tables~\ref{tab:table_GLSErrorP1} and \ref{tab:table_GLSErrorP2}
shows the performance of the method for $p=0.1$ and $p=0.3$. As expected, higher values of $\epsilon$ are able to detect more
errors, but at the cost of increased compressed file size. Table~\ref{tab:table_GLSErrorFinal} shows the efficiency of
the method. Up to $96\%$ of single bit errors introduced at the tail
of the compressed file are detected by a modest increase in the
redundancy (up to $15.78\%$). It should be noted that errors
introduced in the beginning of the compressed file can be very easily
detected by the proposed method.
\subsection{Arithmetic Coding with a Forbidden Symbol: Prior Work}
The idea of using a forbidden symbol into arithmetic coding was
first introduced by Boyd et al.~\cite{Boyd1997}. It was
subsequently studied by Chou et al.~\cite{ARQ},
Grangetto et al.~\cite{GrangettoForbiddenSymbol} and Bi et al.~\cite{TrellisAC}. Thus the idea of using a forbidden symbol that is proposed in this paper is not novel. However, the approach that is taken in this paper is unique and entirely motivated by a non-linear dynamical systems
approach, through the wonderful properties of Cantor set. We are thus able to justify why the method actually works. To the
best of our knowledge, none of the earlier researchers have made
this close connection between error detection/correction for
repetition codes or arithmetic coding and Cantor set. This work paves the way for future research on error correction using fractals/Cantor sets and potentially a host of new efficient techniques using Cantor sets could be designed.
\section{Conclusions and Open Problems}
\label{conclusions}
In this work, we have explored lossless data compression in the presence of noise.
GLS-coding is sensitive to noise because of chaos.  Cantor sets have
paradoxical properties that enable error detection and correction.
Repetition codes are an example of `codewords on a Cantor set' which can
detect and correct errors. By reserving a forbidden symbol on the
interval $[0,1)$, we can ensure that the codewords for GLS-coding
lie on a Cantor set and thereby detect errors while GLS-decoding, and without significantly increasing the compressed file size. This approach can be
applied to any mode of GLS and generalizable to larger alphabets.
This is a way of performing joint source channel coding. However, we
do not know whether other efficient error control codes can be
similarly designed using such Cantor sets (or other fractals in higher dimensions) and whether we can exploit the
structure of the Cantor set to perform efficient error correction.
These are challenging open problems and worth exploring for the future.
\section*{Acknowledgments}
The author express sincere thanks to Prabhakar G Vaidya for introducing him to the fascinating field of Non-linear dynamics/Chaos, Cantor sets and Fractals. The author also express gratitude to William A Pearlman for introducing him to the equally exciting field of data compression.

\end{document}